\documentclass[fleqn,twoside]{article}
\usepackage{espcrc2}

\usepackage{graphicx}
\usepackage[figuresright]{rotating}

\newcommand{\be}{\begin{equation}}
\newcommand{\ee}{\end{equation}}
\newcommand{\bea}{\begin{eqnarray}}
\newcommand{\eea}{\end{eqnarray}}
\newcommand{\bt}{\begin{tabular}}
\newcommand{\et}{\end{tabular}}

\newcommand{\pp}{~~~.}
\newcommand{\vv}{~,}

\newcommand{\etal}{{\it et al.}}
\newcommand{\AmS}{{\protect\the\textfont2
  A\kern-.1667em\lower.5ex\hbox{M}\kern-.125emS}}

\title{Neutrinos and Primordial Nucleosynthesis}

\author{G. Mangano\address[unina]{Dipartimento di Scienze Fisiche, Universit\`{a} di
Napoli {\it Federico II} and INFN, Sezione di Napoli, \\Complesso
Universitario di Monte Sant'Angelo, Via Cintia, I-80126 Napoli,
Italy\\\texttt{mangano@na.infn.it}} and P.D.
Serpico\address[maxplanck]{Max Planck Institut f\"{u}r Physik,
Werner Heisenberg Institut,
\\F\"{o}hringer Ring 6, 80805, M\"{u}nchen, Germany\\\texttt{serpico@mppmu.mpg.de}}}

\begin{document}

\begin{abstract}
The importance of the Big Bang Nucleosynthesis (BBN) as a unique
tool for studying neutrino properties is discussed, and the recent
steps towards a self-consistent and robust handling of the weak
reaction decoupling from the thermal bath as well as of the
neutrino reheating following the $e^{+}-e^{-}$ annihilation are
summarized. We also emphasize the important role of the Cosmic
Microwave Background (CMB) anisotropy in providing an accurate and
independent determination of the baryon density parameter
$\omega_b$. The BBN is presently a powerful parameter-free theory
that can test the standard scenario of the neutrino decoupling in
the early Universe. Moreover it can constrain new physics in the
neutrino sector. The perspectives for improvements in the next
years are outlined. \vspace{1pc}
\end{abstract}

\maketitle

\section{Introduction}
The Standard Cosmological Model predicts the existence of a
neutrino background (C$\nu$B) filling our Universe with densities
of the order $n_\nu\approx 100\,cm^{-3}$ per flavor and thermal
energies of ${\cal O}$(1~K), which the data for the mass splitting
coming from the neutrino oscillation experiments put nowadays in
the non-relativistic regime. As for the much better studied photon
microwave background (CMB), a detailed analysis of the C$\nu$B
properties would provide a unique window on the physical
conditions in the early stages of life of the Universe. Moreover,
the peculiar environment of a \emph{thermalized neutrino medium},
impossible to reproduce in laboratory experiments, may give in
principle some insights on exotic properties of neutrino physics,
as the existence of sterile degrees of freedom that may be excited
in such extreme conditions.

The incredibly weak interaction of the neutrinos, especially at
such low energies, makes hopeless at present any perspective of
direct detection of C$\nu$B. Nevertheless, given their extremely
low interaction rate, the natural out-of-equilibrium driving force
of the expansion of the Universe pushed them to decouple from the
thermal bath much earlier than the CMB, when the temperature was
${\cal O}$(1~MeV). This temperature is close to the electron mass
$m_e$, setting the scale of the electron/positron annihilation,
and both are close to the ${\cal O}$(0.1~MeV) scale of the
synthesis of the light nuclei via thermonuclear fusion. This
suggests that interesting phenomena involving the C$\nu$B indeed
can affect the pattern of nuclides coming from the cosmic
cauldron.

BBN is a privileged laboratory for the C$\nu$B studies with
respect to other cosmological probes, such as CMB anisotropies or
the Large Scale Structure (LSS), since it is sensitive to the
$\nu$ (weak) interactions as well as to the shape of the
$\nu_e-\bar{\nu}_e$ phase space distributions entering the
$n\leftrightarrow p$ inter-conversion rates
\begin{eqnarray}
\nu_e + n &\leftrightarrow& e^- + p \vv \nonumber \\
\bar{\nu}_e + p &\leftrightarrow& e^+ + n \vv \nonumber \\
n &\leftrightarrow& e^- + \bar{\nu}_e + p \pp \label{e:reaction}
\end{eqnarray}
Apart from the energy density due to the extra ({\it i.e.} non
electromagnetic) relativistic degrees of freedom, typically
parameterized via an effective number of neutrinos $N_{\rm eff}$,
the BBN tests the \emph{dynamical} properties of the neutrinos in
a thermalized (almost) CP-symmetric medium.

Other cosmological observables are instead sensitive only to the
C$\nu$B \emph{gravitational} interaction. It follows that CMB
mainly probes $N_{\rm eff}$ at a much later epoch and the LSS
(mainly) the neutrino mass scale $m_\nu$, since neutrinos turn
into non-relativistic species just in time to influence the
structure formation dynamics.

Still few years ago, the BBN theory together with the observations
of the abundances of pristine nuclides were used to determine the
baryon to photon ratio $\eta\equiv n_B/n_\gamma$ or, equivalently,
the baryon fraction of the universe $\omega_b= \Omega_b h^2$.
Nowadays $\omega_b\approx 0.023$ is fixed to better than 5\%
accuracy by detailed CMB anisotropies
analysis~\cite{Spergel:2003cb}, thus leaving the BBN as an
over-constrained (and thus, very predictive) theory. Once
$\omega_b=0.023 \pm 0.001$ is plugged into the BBN theory, the
prediction for the deuterium, which is the nuclide most sensitive
to $\omega_b$, nicely fits the range of the observed values in
high redshift, damped Ly-$\alpha$ QSO
systems~\cite{Kirkman:2003uv}, thus offering a remarkable example
of internal consistency of the current cosmological scenario.
Moreover, the predictions of other light nuclei which at least
qualitatively agree with the observed values are likely to put
constraints on the Galactic chemical evolution ($^3$He) or to the
temperature scale calibration or depletion mechanisms in PopII
halo stars ($^7$Li).

Apart from the uncertainty on $\omega_b$, the $^2$H, $^3$He and
$^7$Li abundance predictions are mainly affected by the nuclear
reaction uncertainties. An updated and critical review of the
nuclear network and a new protocol to perform the nuclear data
regression has been presented in~\cite{Serpico:2004ba} and widely
discussed in~\cite{Serpico:2004gx}, to which we address for
details.

On the other hand, the predicted value of the $^4$He mass
abundance, $Y_p$, is poorly sensitive to the nuclear network
details and has only a weak, logarithmic dependence on $\omega_b$,
being fixed essentially by the ratio of neutron to proton number
density at the onset of nucleosynthesis. Its crucial dependency on
the weak rates (\ref{e:reaction}) and on the (standard or exotic)
neutrino properties will be briefly discussed in the following
section.

\section{Weak Rates, $^4$He, and relic neutrinos}
As a first approximation, the neutrino decoupling can be described
as an instantaneous process taking place around 2-4 MeV, without
any overlap in time with $e^{+}-e^{-}$ annihilation. All $\nu$
species would then keep perfect Fermi-Dirac distributions, with
temperature $T_\nu$ smaller than the photon one $T$ since they do
not benefit in this instantaneous decoupling scheme of the entropy
release from $e^{+}-e^{-}$ annihilations. The asymptotic ratio
$T/T_\nu$ for $T<<m_e$ can be evaluated in an analytic way, and
turns out to be $(11/4)^{1/3}\approx 1.401$.

\begin{figure}[htb]
\includegraphics[width=6.0cm]{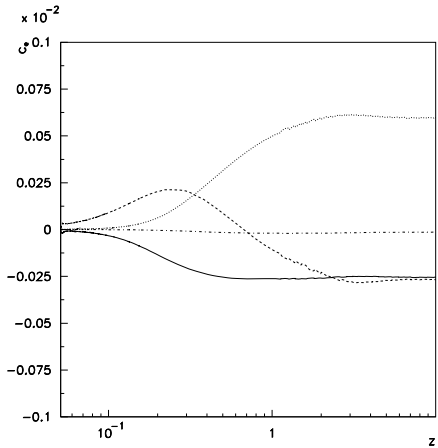}
\caption{The evolution of the electron neutrino distortion
coefficients $c_0^e$ (solid), $c_1^e$ (dashed), $c_2^e$ (dotted)
and $c_3^e$ (dot-dashed) versus $z= m_e/T$.}\label{fig:nuecoef}
\end{figure}

\begin{figure}[htb]
\includegraphics[width=6.0cm]{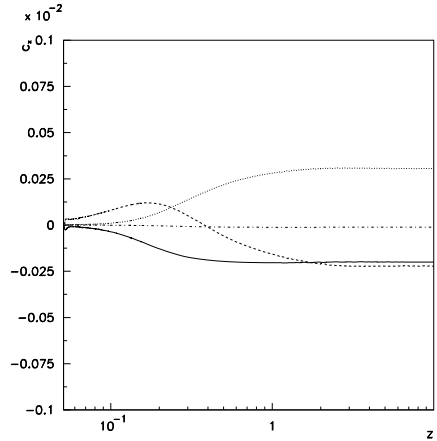}
\caption{The evolution of the $\mu$ and $\tau$ (collectively
denoted by $x$) neutrino distortion coefficients $c_0^x$ (solid),
$c_1^x$ (dashed), $c_2^x$ (dotted) and $c_3^x$ (dot-dashed) versus
$z= m_e/T$.}\label{fig:nuxcoef}
\end{figure}

\begin{figure}[htb]
\includegraphics[width=6.0cm]{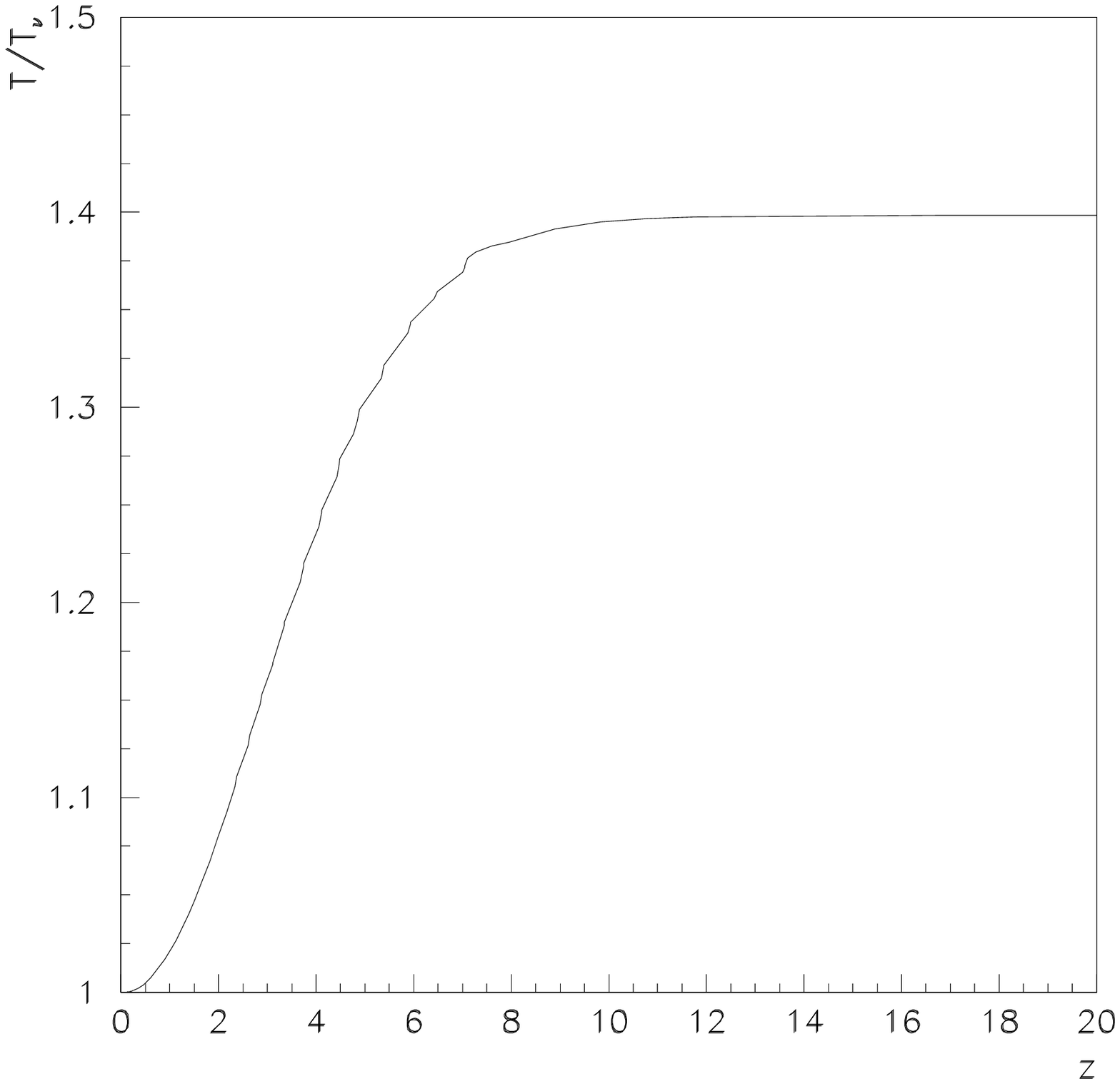}
\caption{The evolution of $T/T_\nu$ versus $z= m_e/T$. The
asymptotic value at small temperatures is 1.3984.}
\label{fig:zbar}
\end{figure}

More accurate calculations by solving the kinetic equations have
been performed, and they show a partial entropy transfer to the
neutrino plasma. As a consequence, the neutrino distributions get
distorted. In \cite{Mangano:2001iu,Esposito:2000hi} it was shown
that with a very good approximation the distortion in the
$\alpha$-th flavor can be described as
\begin{equation}
f_{\nu_\alpha} \left( x,y\right) \simeq \frac{1}{e^{y }+1} \left(
1 + \sum_{i=0}^3\, c_{i}^\alpha(x)\, y^i\right) \, \pp
\label{expan2}
\end{equation}
where $x\equiv m_e/T_\nu$, and $ y\equiv p/T_\nu$. The evolution
of the $c_{i}^e$ and $c_i^x$ with $x=\mu$,$\tau$ are shown in
Figure \ref{fig:nuecoef} and \ref{fig:nuxcoef} versus $z= m_e/T$,
respectively.
Notice that the electron neutrinos get a larger entropy transfer
than the $\mu$ and $\tau$ since they also interact via charged
currents with the $e^{\pm}$ plasma. By fully consistently
including order $\alpha$ QED corrections to the photon and
$e^{\pm}$ equation of state, in~\cite{Mangano:2001iu} the energy
density in the neutrino fluid is found to be enhanced by 0.935\%
(for $\nu_e$) and 0.390\% (for $\nu_\mu$ and $\nu_\tau$) and the
effective ratio $T/T_\nu\approx 1.3984$ is slightly lower than the
previous instantaneous decoupling estimate. In Figure
\ref{fig:zbar} we show the evolution of this ratio versus $z$. Put
in terms of $N_{\rm eff}$, the standard prediction is then 3.04
instead of 3.

\begin{figure}[htb]
\includegraphics[width=6.0cm]{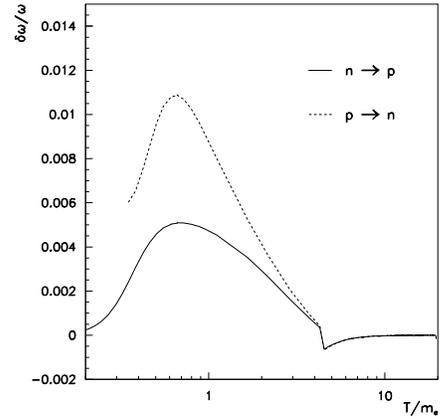}
\caption{The relative correction to the $n \rightarrow p$ (solid
line) and $p \rightarrow n$ (dashed line) total rates, due to the
neutrino distortion. $\delta \omega$ is the difference of the $n
\leftrightarrow p$ rates calculated by numerically solving the
neutrino decoupling and the ones for an instantaneous neutrino
decoupling at $T_D=2.3$ MeV. For $p \rightarrow n$ the effect is
shown down to $T \sim 0.3$ MeV, since for lower temperatures the
total rate becomes negligibly small.} \label{fig:newcorrnu}
\end{figure}

How much the $^4$He prediction is influenced by these phenomena?
In several papers (see \cite{lopez,EMMP} and references therein)
the value of $Y_p$ has been computed by improving the evaluation
of the rates (\ref{e:reaction}) including electromagnetic
radiative corrections, finite mass corrections and weak magnetism
effects, as well as the plasma and thermal radiative effects. In
particular in \cite{Serpico:2004gx} it has been also considered
the effect of the neutrino spectra distortions and of the process
\begin{equation}
~\gamma+p \leftrightarrow \nu_e + e^{+}+ n,
\end{equation}
which is kinematically forbidden in vacuum, but allowed in the
thermal bath. The latter is shown to give a negligible
contribution, while the neutrino distortion has a significant
influence on the rates only at relatively late times (see Figure
\ref{fig:newcorrnu}), when however the neutron to proton density
ratio is already frozen. Nonetheless, one would expect effects up
to ${\cal O}$(1\%) due to the neutrino reheating. However, the
spectral distortion and the changes in the energy density and
$T_\nu(T)$ conspire to almost cancel each other, so that $Y_p$ is
changed by a sub-leading ${\cal O}$(0.1\%). This effect is of the
same order of the predicted uncertainty coming from the error on
the measured neutron lifetime, $\tau_n=885.7\pm 0.8$ s~\cite{PDG},
and has to be included in quoting the theoretical prediction,
$Y_p=0.2481\pm 0.0004\:(1\,\sigma$, for $\omega_b=0.023$).

Unfortunately, there is no hope at present to single out such a
small effect, given the observational situation plagued by
systematics. The value of $Y_p$ is usually derived by
extrapolating to zero metallicity the measurements done in dwarf
irregular and blue compact galaxies, that are among the least
chemically evolved galaxies. The typical statistical errors are of
the order of 0.002 ({\it i.e.}, at the 1\% level), but the
systematics are such that in the recent reanalysis
\cite{Olive:2004kq} the authors argue for the conservative range
$0.232\leq Y_p\leq 0.258$, {\it i.e.} a 1 $\sigma$ error of ${\cal
O}$(5\%) . On the other hand, it is worth stressing that BBN
provides the best limit on some C$\nu$B properties. For example,
the range $0.232\leq Y_p\leq 0.258$ gives the limit $-1.2\leq
\Delta N_{\rm eff}\leq 0.7$, only slightly dependent on
$\omega_b$, and much stronger than the typical actual CMB
constraint on this parameter.

Similarly, a possible C$\nu$B asymmetry due to non-vanishing
neutrino chemical potentials is much better constrained by the
BBN, which limits (see {\it e.g.} \cite{Cuoco:2003cu}) \be
\left|\xi_{e}\right|\equiv \left|\mu_{\nu_e}/T_\nu\right|\leq
{\cal O}(0.1),\label{xibound}\ee while both CMB or LSS probes are
only sensitive to $\xi\sim$ ${\cal O}$(1). The former is indeed
mainly fixed by the effect of the (slightly degenerate)
$\nu_e-\bar{\nu}_e$ distributions in the weak rates
(\ref{e:reaction}), while the latter are only sensitive to the
extra energy density present in the $\xi\neq 0$ case. It was
recently realized~\cite{Dolgov:2002ab} that because of flavor
oscillation in the primordial plasma, using present determination
of mass differences and mixing angles from atmospheric and solar
neutrinos, the three asymmetry parameters $\xi_\alpha$ should be
very close each other, so the bound of Equation (\ref{xibound})
can be extended to all neutrino flavors.

In conclusion, the precision cosmology era has opened a new
opportunity for the BBN theory to probe the very early universe,
and in particular neutrino physics in such a unique environment.
The theoretical predictions are in this respect quite robust, and
will be further sharpened by new nuclear reactions measurements,
as well as by refining the CMB determination of $\omega_b$,
especially for the $^2$H, $^3$He and $^7$Li nuclides. For the
$^4$He, which is the most sensitive probe of new physics, only a
significant improvement in understanding the  observational
systematics could offer deeper insights on physics of the early
Universe.

\end{document}